\documentclass[journal]{IEEEtran}
\usepackage{amssymb,amsfonts}
\usepackage{algorithmic}
\usepackage[ruled, noend]{algorithm2e}
\usepackage{enumerate}
\usepackage{hyperref,xcolor}

\usepackage{tabularx}
\usepackage{booktabs}
\usepackage{multirow}

\usepackage{cite}

\usepackage{graphicx}
\ifCLASSINFOpdf
\else
\fi
\usepackage{amsmath}
\usepackage{array}
\usepackage{url}

\UseRawInputEncoding

\begin{document}
%
\title{TENSILE: A Tensor granularity dynamic GPU memory scheduling method toward multiple dynamic workloads system}
%
%
%

\author{Kaixin Zhang, 
        Hongzhi Wang\thanks{Hongzhi Wang is corresponding author.},
        Han Hu,
        Songling Zou,
        Jiye Qiu,
        Tongxin Li,
        Zhishun Wang
        }

%
%

\markboth{Journal of \LaTeX\ Class Files,~Vol.~14, No.~8, August~2015}%
{Shell \MakeLowercase{\textit{et al.}}: Bare Demo of IEEEtran.cls for IEEE Journals}
%



This work has been submitted to the IEEE for possible publication. Copyright may be transferred without notice, after which this version may no longer be accessible.
\newpage

\maketitle

\begin{abstract}
Recently, deep learning has been an area of intense research. However, as a kind of computing-intensive task, deep learning highly relies on the scale of GPU memory, which is usually prohibitive and scarce. Although some extensive works have been proposed for dynamic GPU memory management, they are hard to apply to systems with multiple dynamic workloads, such as in-database machine learning systems.

In this paper, we demonstrated TENSILE, a method of managing GPU memory in tensor granularity to reduce the GPU memory peak, considering the multiple dynamic workloads. TENSILE tackled the cold-starting and across-iteration scheduling problem existing in previous works. We implemented TENSILE on a deep learning framework built by ourselves and evaluated its performance. The experiment results show that TENSILE can save more GPU memory with less extra overhead than prior works in single and multiple dynamic workloads scenarios.
\end{abstract}


%
\IEEEpeerreviewmaketitle

\section{Introduction}
\label{Introduction}
\IEEEPARstart{D}{eep} learning is widely used in many areas of data analysis. With appropriate training, deep learning models can achieve much higher performance than traditional ones \cite{DLSurvey}. However, a significant problem of deep learning is that it is usually costly for GPU memory, especially in the communities of image recognition and natural language processing, whose models like ResNet \cite{ResNet}, BERT \cite{BERT}, and GPT-3 \cite{GPT-3} have as many as billions of parameters. These parameters and the corresponding feature maps cause massive GPU memory footprint \cite{Efficient}, which impedes their application. For example, such deep neural networks(DNN) could hardly be performed in a database since building a large GPU cluster is prohibitive for the existing database systems. Thus, the increasing size of deep learning models acquires that the GPU memory must be managed more efficiently. 

Additionally, many scenarios of deep learning have multiple dynamic workloads, such as cloud database services \cite{Narasayya2015SharingBP}, in-database machine learning \cite{Li2018EasemlTM, Karlas2018EasemlIA}, cloud deep learning services \cite{Multi-tenant_Deeplearning, GPU_Cluster, Narayanan2020HeterogeneityAwareCS}, and AI-driven DBMSes \cite{Neo, Bao, LearnedIndex}. Although the memory occupied by a single load is limited, the total amount of running jobs is huge, so there needs to be enough GPU memory for each job. For example, prior work \cite{Narayanan2020HeterogeneityAwareCS} mentioned that there are situations in that jobs cannot be co-located due to the GPU memory constraints in cluster scheduling problems. To tackle such problems under multiple dynamic workload scenarios, we propose TENSILE to reduce the total GPU memory peak.

The \textbf{critical problem} is how to reduce the GPU memory peak of each and entire computation process efficiently and update the scheduling plan in time to fit the dynamic workloads. Although there are already some efficient methods \cite{Capuchin, vDNN, SuperNeurons} to reduce the GPU memory cost by swapping and recomputation, the following three issues are still unsolved. 

\begin{figure*}[h]
\centerline{\includegraphics[width=\linewidth]{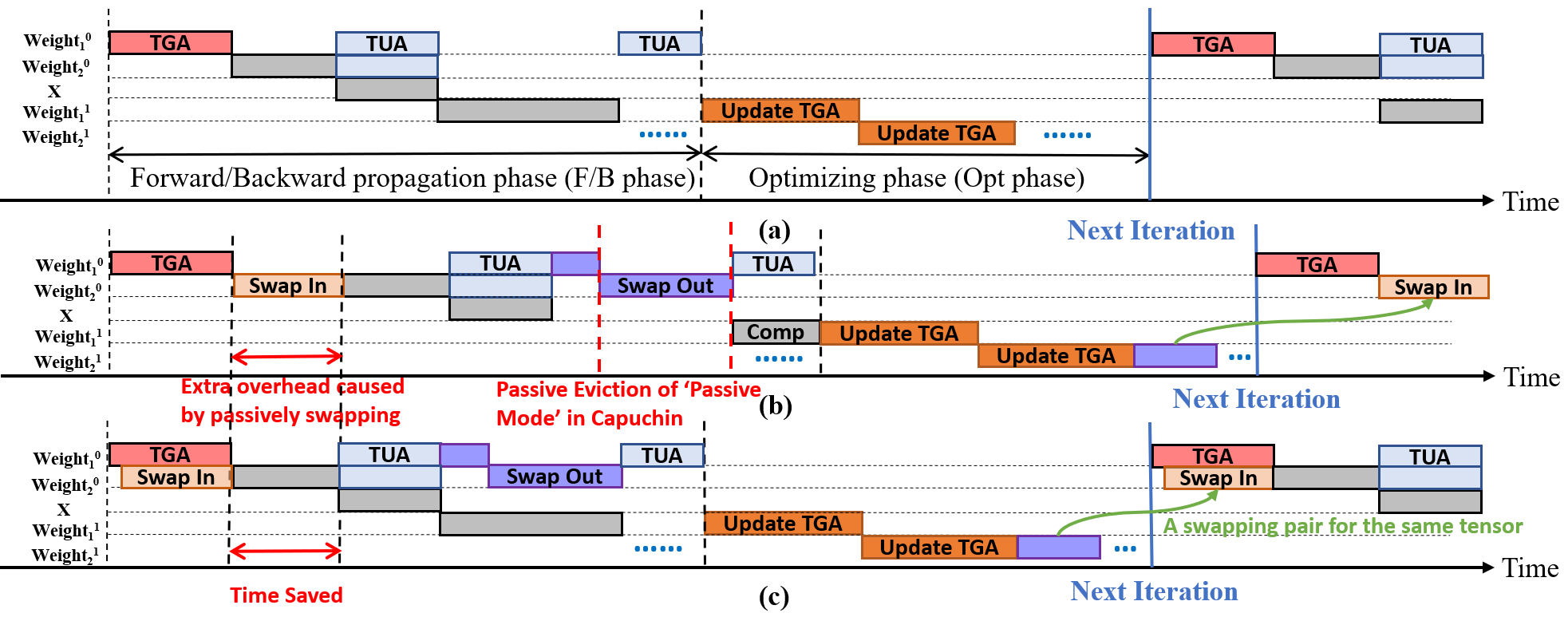}}
\caption{Tensor Access and Memory peak, $weight_1^i$ means the $weight_1$'s value in the $i^{th}$ iteration, and the update \textit{TUA} means the operator of updating the weights. (a) is the situation without scheduling; (b) is the scheduling that cannot across iterations, such scheduling will cause extra time overhead. It also shows the cold-starting problem of Capuchin; (c) is our method's effect of reducing GPU memory peak}
\label{fig:Across Iteration}
\vspace{-0.2cm}
\end{figure*}

\begin{itemize}
    \item \textbf{Multiple dynamic workloads.} To our best knowledge, most existing approaches are designed to schedule GPU memory for a single workload. When it comes to scenarios with multiple dynamic workloads, these methods can hardly manage the GPU memory as well as they were on a single job. Scheduling GPU memory with multiple dynamic workloads is challenging since the workloads are executed asynchronously, and such a mechanism will cause dynamic tensor access patterns. The core challenge is to update the scheduling plan with its fluctuation. Also, choosing the proper tensors that need to be scheduled is critical to scheduling efficiency. Furthermore, the challenge involves the design of the system architecture to support such kinds of scheduling algorithms.
   
    \item \textbf{Cold-starting.} The most advanced method, such as Capuchin \cite{Capuchin} which schedules in tensor granularity, needs to observe the tensor access pattern first before generating the scheduling plan. As shown in \autoref{fig:Across Iteration}(b), such 'Passive Mode' will cause extra overhead by passively swapping tensors when the GPU memory is insufficient. Also, the multiple workload scenario causes the observation to be inaccurate. So the core challenge is to find a proper way to initialize the tensor access pattern.
    
    \item \textbf{Across-iteration Scheduling.} All prior methods only schedule within a single iteration. They ignore that scheduling across iterations is also necessary since some tensors evicted in the current iteration's Opt phase are used as inputs in the next iteration’s F/B phase, as shown in \autoref{fig:Across Iteration}, which leads to extra overhead. The core challenge is to consider the swapping between two phases.
\end{itemize}

In this paper, we tackle these challenges and propose TENSILE, a run-time dynamic GPU memory scheduling method of tensor granularity designed to reduce the GPU memory footprint peak towards multiple dynamic workloads.

For the first problem of multi-workloads, we develop a scheduling method that can keep tracking the tensor access latency for updating the scheduling plan to support the multiple dynamic workloads. By updating the scheduling plan in time, the dynamic tensor access patterns caused by multi-workloads scenarios can be well adapted. 

To solve the second problem, we use a light neural network model to predict the latency of each operator under current GPU usage. With this model, TENSILE can avoid the extra overhead caused by the 'Passive Mode'. 

And for the third problem, our method could schedule tensors across iterations. TENSILE can swap out the parameters at the Opt phase and prefetch them before they are used in the following computation iteration. As the schematic \autoref{fig:Across Iteration}(b, c) shows, such an approach can effectively reduce the total GPU memory peak with less extra overhead of passively swapping and supporting multiple dynamic workloads.

TENSILE is a tensor granularity scheduling approach for supporting deep learning tasks based on tensor computation, such as the training of CNN \cite{CNN} and GCN \cite{GCN}. Additionally, we can achieve higher performance with scheduling in tensor granularity since tensor is the most fundamental component of deep learning algorithms. Also, benefiting from the tensor granularity, TENSILE is not limited to the scope of application with forward/backward propagation like most prior works do. It can also handle any tensor computation tasks expressed with tensor operators in a static compute graph.

Our contributions can be concluded as:
\begin{itemize}
    \item Inspired by existing GPU memory scheduling methods, we proposed a scheduling method toward multiple dynamic workloads. To support such scheduling, we designed a system to track the jobs' running and manage their memory as a whole, rather than optimizing them separately as previous works did. This method can also schedule the tensors across iterations and reduce the entire process's GPU memory peak to increase the system's model capacity. 
    \item We also proposed an innovative algorithm to generate the swapping and recomputation plan based on tensor access sequence, which can support the multiple dynamic workloads better than prior works. With this algorithm, we could initialize the scheduling plan without measuring first and update the scheduling plan in time with the fluctuation of GPU usage. 
    \item According to the experiment results, TENSILE can achieve a higher GPU memory saving rate with less performance loss, especially in multiple dynamic workload scenarios.
\end{itemize}

\section{Related Work}
\label{Related Work}
In this section, we introduce some prior works in GPU memory management.

NVIDIA is the first organization to pay attention to the GPU management problem. It integrated the Unified Memory technology in CUDA6\footnote{https://developer.nvidia.com/blog/unified-memory-in-cuda-6/}. This work allows programmers to use the GPU and Host Memory as a unified space. Unlike TENSILE, this work is not specially designed to schedule deep learning processes, which implies programmers need to decide when and which tensor should be swapped with CUDA programming.

In the area of GPU memory scheduling for deep learning, Minsoo Rhu et al. proposed the vDNN \cite{vDNN}, which is the first work focusing on scheduling the GPU memory footprint for deep learning processes. The central part of vDNN is an algorithm to decide which layer's interim result should be released, evicted, or prefetched. Although it is a pioneering work, some problems remain, such as delayed computation start, high pinned memory requirements, and GPU memory fragmentation. Shriram S.B. et al. improved vDNN and solved the three problems mentioned above \cite{ShriramS2019DynamicMM}. By allowing to overlap of the tensor eviction and prefetching process with the computation process, Shriram S.B. et al. significantly improved the vDNN's performance. Linnan Wang et al. proposed SuperNeurons \cite{SuperNeurons} which is the first dynamic GPU memory run-time management approach. It also introduced recomputation into the scheduling. The authors believe the convolution layer is the most valuable layer to be swapped and designed some approaches to schedule the convolution layers. Although this design helps it to achieve significant efficiency on convolutional networks, it also limits the compatibility for more advanced network structures as transformer \cite{AttentionIsAllYouNeed}. Donglin Yang et al. proposed the first method that supports a unified memory pool for non-linear networks \cite{DongLin_Efficient}. The method can construct the execution order for non-linear networks based on graph analysis so that it could support non-linear networks better than previous works.

Although all the methods above have significant advantages, they have a major shortcoming compared to TENSILE: all of them are designed to schedule GPU memory in layer granularity. Such a mechanism makes them lack compatibility and unable to save GPU memory as much as tensor granularity methods like TENSILE. This is the fundamental reason why they can not schedule the newest deep learning models that contain complex structures inside a 'layer' (or as known as 'block'), such as transformer-based networks \cite{AttentionIsAllYouNeed} and the Capsule Network \cite{CapusleNetwork}. And the tensors inside the 'layer' can not be scheduled by these methods, and neither do the interim tensors in the optimizer, such as the $1^{st}\ moment\ vector$ and the $2^{nd}\ moment\ vector$ in the Adam optimizer \cite{Adam}. These works aim to improve the batch size in a single workload scenario, so only scheduling the tensors in the F/B phase is enough. However, we aim to reduce the memory footprint peak to make the system support more job training simultaneously when it comes to multiple workload scenarios. Moreover, such methods are not good enough since the GPU memory peak will be not only appear in the F/B phase but also in the Opt phase.


In 2019, another method was proposed \cite{Zhang2019EfficientMM}, the first approach that can schedule GPU memory in tensor granularity. This method proposed two orthogonal approaches to reduce the memory cost from the system perspective. It reduces the GPU memory peak by managing the memory pool and scheduling the swapping with a heuristic method. Benefiting from its tensor granularity, this method can support any deep learning network. Xuan Peng et al. proposed Capuchin \cite{Capuchin}. Capuchin can observe the tensor access pattern and decide which tensor to be swapped and which to be recomputed with a value-based approach. This method is more flexible in supporting various types of neural networks than those layer granularity methods. However, it highly depends on observing the tensor access pattern when the calculation is running for the first time since the tensor access pattern is needed to generate the scheduling plan. Such observation leads to passive swapping and PCIe channel occupancy when the GPU memory is insufficient.
\begin{figure}[h]
  \centering
  \includegraphics[width=\linewidth]{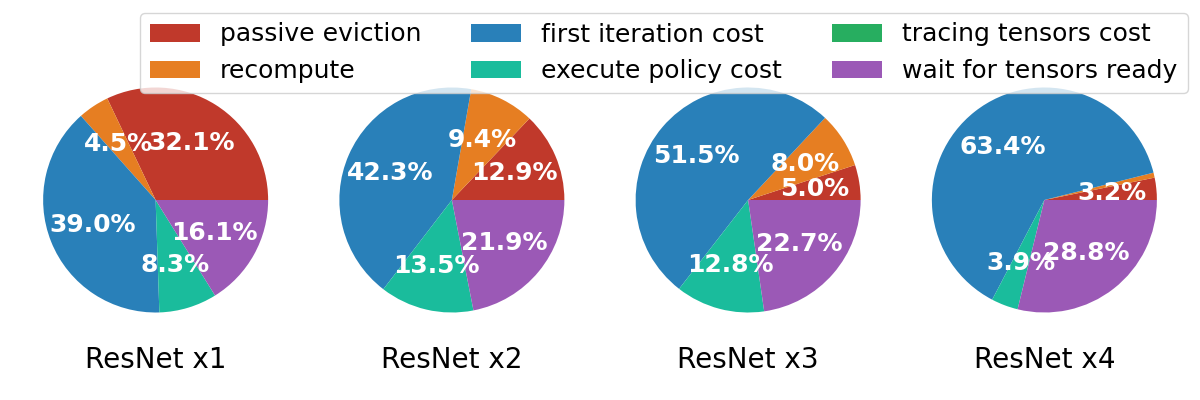}
  \caption{Overhead Analysis of Capuchin. The 'first iteration cost' corresponds to the 'Passive Mode' of Capuchin, and the 'wait for tensors ready' corresponds to the effectiveness of swapping in. All the items except 'first iteration cost' are measured after the first iteration.}
  \label{fig:Overhead Analysis of Capuchin}
\end{figure}

These two advanced works schedule GPU memory in tensor granularity and solve compatibility problems. However, compared to TENSILE, neither of these prior works can solve the three problems we propose in Section I. We take Capuchin as an example to explain it. 

In an environment with multiple workloads, influenced by multiple workloads, the GPU usage fluctuates irregularly, which causes the operators' latency and tensor access pattern to keep changing, so the scheduling plan must be able to be kept updating iteratively. Capuchin makes a fixed scheduling plan once at the first epoch based on the observed tensor access pattern in the 'Passive Mode', which is not feasible because the latency of the operator varies with the dynamic load of multiple jobs. Especially when a job launches, the GPU usage will change drastically and cause the initial scheduling plan to lose efficacy. Capuchin has not enough ability to adjust the plan to fit the fluctuation of the tensor access pattern in multiple dynamic workload scenarios. This can be proved by \autoref{fig:Overhead Analysis of Capuchin}, the time cost of 'wait for tensors ready' increases with the number of workloads.  

The 'Passive Mode' also causes the cold-starting discussed problem in \autoref{Introduction}. According to the 'first iteration cost' of \autoref{fig:Overhead Analysis of Capuchin}, this unexpected eviction causes the time cost of the first iteration increases since passive swap-out causes the computation to block.
 
Also, Capuchin and other works above only considered the tensor access in a single iteration. For example, the updated parameters' accesses at the end of the iteration can not be swapped proactively. Thus, these parameters can only be swapped passively at the next iteration with an additional time overhead. As shown in \autoref{fig:Overhead Analysis of Capuchin}, the passive eviction takes 32.1\% of the overall time cost. Most of these evictions are caused by the memory peak caused by the unexpected swap-in across iterations.

\section{Overview}
This section defines the GPU memory scheduling problem and overviews our deep learning system built to support TENSILE.

\subsection{Preliminary}
For the convenience of scheduling, we use a graph model called \emph{static compute graph} \cite{tensorflow} to describe the tensor calculation jobs. Let $V$ be the set of all basic operators to manipulate tensors, $E$ be a set of tensors generated by operators in $V$, and $S$ be a set of scheduling events, whose elements are \{Swap-out, Swap-in, Recompute, Release\}. With these symbols, we define the multiple dynamic workload scheduling problem.

\begin{figure}[h]
  \centering
  \includegraphics[width=\linewidth]{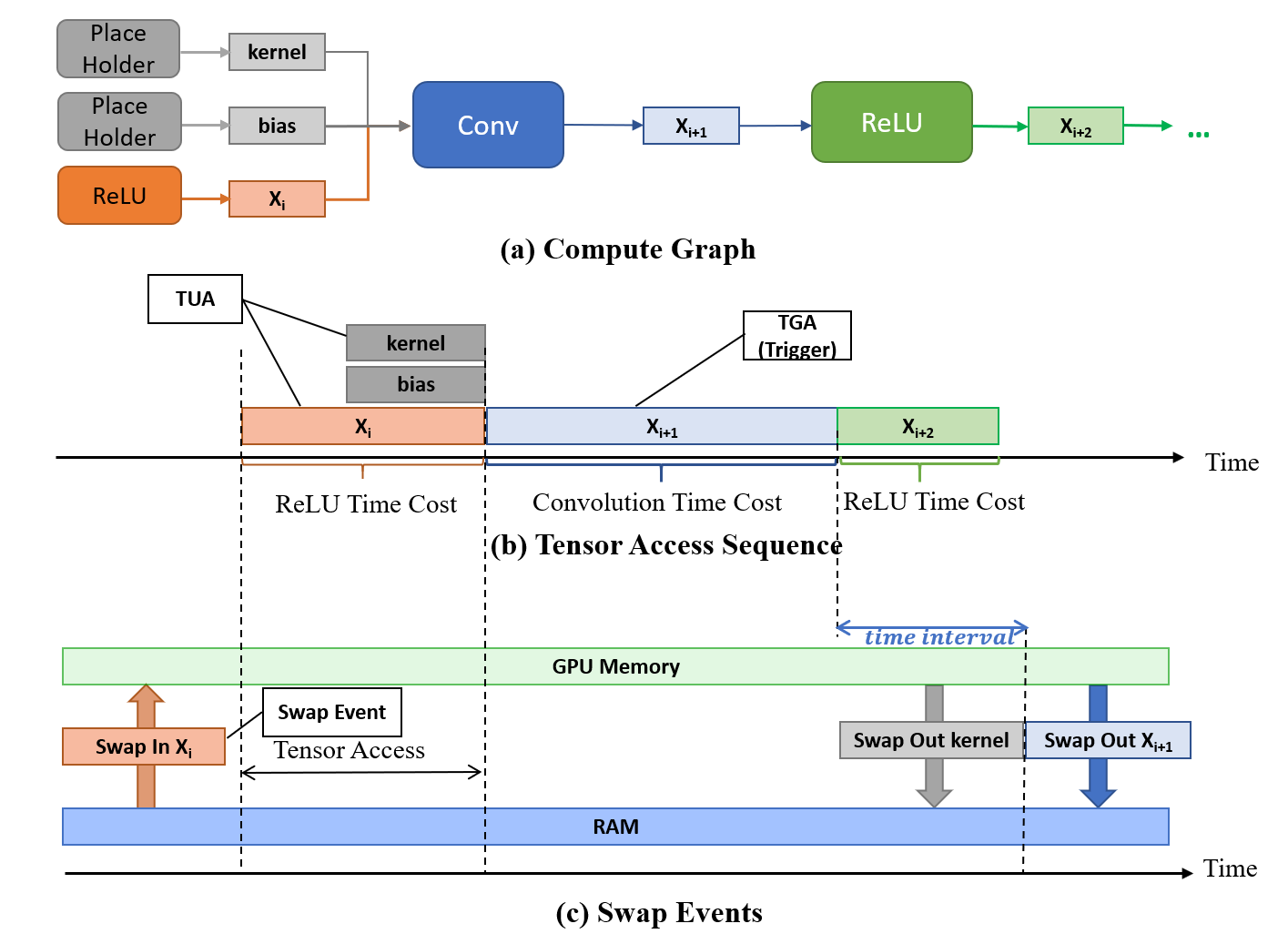}
  \caption{Demonstration of the Concepts, (a) is a simple compute graph of CNN; (b) is a tensor access sequence corresponding to (a); (c) is a scheduling plan of (b).}
  \label{fig:Concepts' Examples}
  \vspace{-0.5cm}
\end{figure}

\textit{\textbf{Job:}} A deep learning job $j$ is denoted by a static compute graph as $G_j(V,E)$. Note that the graph is a DAG (Direct Acyclic Graph) just like the one in TensorFlow \cite{tensorflow}, and the operators of the optimizer are also included in $G_j(V,E)$. An example of the graph is shown in \autoref{fig:Concepts' Examples}(a), a convolution operator is represented as a node in the graph. Such a node takes multiple tensors as the input and outputs other tensors. Such a model could effectively represent the computation in modern deep learning frameworks, such as TensorFlow \cite{tensorflow}. 

\textit{\textbf{Tensor Access:}} For a tensor $t$ in $E$, which is generated by the operator $\mu$ and used by the operator $\upsilon$, we define the execution of $\mu$ as the \textbf{Tensor Generating Access(TGA)} of $t$ and the execution of $\upsilon$ is the \textbf{Tensor Using Access(TUA)} of $t$. Both tensor accesses are denoted as $a_{j}^{i}$, where $i$ is the access index. Usually, if a given operator $\mu$ is not at the start or the end of a job, it corresponds to two kinds of tensor accesses, the \textit{TUA} and the \textit{TGA}. For example, as shown in \autoref{fig:Concepts' Examples}(b), the convolution operator takes tensors kernel, bias, and $x$ generated by the previous operator as inputs and outputs the feature map tensor $y$. The latency of a \textit{Tensor Access} is defined as the corresponding operator's execution time cost, which can be inferred from the compute graph or logged at run-time. The \textit{TUA} is access of the input tensors of $\mu$, and the \textit{TGA} represents the accesses of the output tensors of $\mu$. Additionally, since the operator \textit{placeholder} generates the input tensors of the whole compute graph, we regard the \textit{placeholder} as a \textit{TGA}. 

\textit{\textbf{Workload:}} A workload $W_j$ is denoted by a \textbf{tensor access sequence} $<a_{j}^{1}$,$a_{j}^{2},...>$, which is ordered by the topological order of job $G_j(V,E)$. The topological order corresponds to the execution order of operators in $V$. When a workload $W_j$ is launched, the framework executes the operators of $W_j$ in topological order. A \textbf{dynamic workload} has tensor accesses whose latency keeps changing at run time.

\textit{\textbf{Scheduling Plan:}} We denote a scheduling plan for workload $W_j$ as a sequence $S_j$ of $s_{j}^{i,t}$, such as $<s_{j}^{i_0,t_0}, s_{j}^{i_1,t_1}, ...>$. And for each event $s_{j}^{i,t}$, $j$ represents that it belongs to the scheduling plan of workload $W_j$, $i$ is its trigger tensor access, which is the prior tensor access of the scheduled event, and $t$ is the time interval between the execution end of the trigger tensor access and the start of the scheduling event, just as \autoref{fig:Concepts' Examples}(c) shows


\textit{\textbf{Memory Footprint:}} Let function $Mem(\cdot)$ be the GPU memory changing caused by tensor access $a$ or scheduling event $x$, the \textbf{memory footprint} of a workload $W_j$ at the end of operator k is 
\begin{equation}
    \setlength{\jot}{0pt}
    \label{equation_MFkj}
    \begin{aligned}
       MF_{j}^{k}=\sum_{i=0}^{k}Mem(a_{j}^{i})+Mem(x_{j}^{i,t})
    \end{aligned}
\end{equation}
, and the \textbf{memory footprint peak} of workload $W_j$ (with $m$ operators) is 
\begin{equation}
    \setlength{\jot}{0pt}
    \label{equation_MFj}
    \begin{aligned}
       MP_{j}=\max \limits_{0\le k\le m}MF_{j}^{k}=\max \limits_{0\le k\le m}\sum_{i=0}^{k}Mem(a_{j}^{i})+Mem(x_{j}^{i,t})
    \end{aligned}
\end{equation}
The summation process in \autoref{equation_MFkj} can be regarded as a discrete GPU memory occupancy change process as the workload's tensor access sequence is executed. Assuming there are $n$ workloads running asynchronously in the system, the \textbf{global memory footprint peak} is $MP=\sum_{j=0}^{n}MP_{j}$. 

\subsection{Problem Definition}
With these symbols above, we define the problem here. Let $M$ be the GPU memory size, our \textbf{multiple dynamic workloads scheduling problem} is how to generate a scheduling plan $X_j$ for each workload $W_j$ to minimize the global memory footprint peak $MP$ with as little extra overhead. So that the system with GPU memory $M$ can run more neural networks simultaneously or run these given workloads with larger batch sizes for higher efficiency. The optimization goal is formalized as \autoref{equation_goal} shows.
\begin{equation}
    \setlength{\jot}{0pt}
    \label{equation_goal}
    \begin{aligned}
        J(X) = \mathop{\arg\min}_{\{X_0,...X_j\}}\sum_{j=0}^{n}\max \limits_{0\le i\le m}\sum_{i=0}^{k}(Mem(a_{j}^{i})+Mem(x_{j}^{i,t}))
    \end{aligned}
\end{equation}

Note that the interim results and the operators of the optimizer are also included in $G$, and can be scheduled like other tensors, which is different from prior works. Those works are only designed to save GPU memory during the forward/backward propagation to achieve a larger batch size or deeper network. However, our method is designed toward multiple dynamic workloads in the system. The method must reduce the entire job's GPU memory peak to support multiple workloads, including the peak caused by the optimizer. Otherwise, when two peaks caused by different jobs coincide, an OOM (Out Of Memory) exception will be triggered. Although the exception can be handled by passive swapping just like the Capuchin \cite{Capuchin} does, it will cause much extra overhead. This feature is practical, especially when using an optimizer like Adam \cite{Adam}, which will use additional memory twice the size of the parameters.

\subsection{Concepts Explanation}
To facilitate understanding, we explain the basic concepts of TENSILE.

The \textbf{\textit{Tensor Access Sequence}} is the alias of workload. For each tensor, its \textit{Tensor Access Sequence} begins with an \textit{TGA} which generates the tensor and followed by a series of \textit{TUA} which represents the tensor used by operators as input, as shown in \autoref{fig:Across Iteration}.
    
A \textbf{\textit{Swap Event}} transfers data between the GPU and the Host Memory. It has two kinds of types, \textbf{\textit{Swap-Out Event}} and \textbf{\textit{Swap-In Event}}. As \autoref{fig:Concepts' Examples}(c) shows, \textit{Swap-Out Event} of a tensor will copy the tensor from the GPU memory to the Host Memory and then releases it from the GPU memory as soon as it is not occupied, which can be regarded as eviction. A \textit{Swap-In Event} will copy a tensor from the Host Memory to the GPU memory, which is used to prefetch the tensor. It is necessary to emphasize that swap-in does not mean the tensor in the Host Memory will be released. On the contrary, it will remain in the Host Memory until the last access of the tensor is finished, which could save much cost of swap-out.
    
A \textbf{\textit{Recomputation Event}} regenerates the tensors which have already been released from the GPU memory. For a \textit{Tensor Access $T_i$}, we can release it after accessing and recompute it before the \textit{Tensor Access $T_j$} that takes the tensor $i$ as input. 

A \textbf{\textit{Release Event}} releases the corresponding tensor from the GPU memory as soon as it has not been used. 

With these concepts, we can offload tensors when not used and then prefetch or regenerate them by swapping or recomputing them before their \textit{TUA}. 


\subsection{System Architecture}
\begin{figure}[h]
  \centering
  \includegraphics[width=\linewidth]{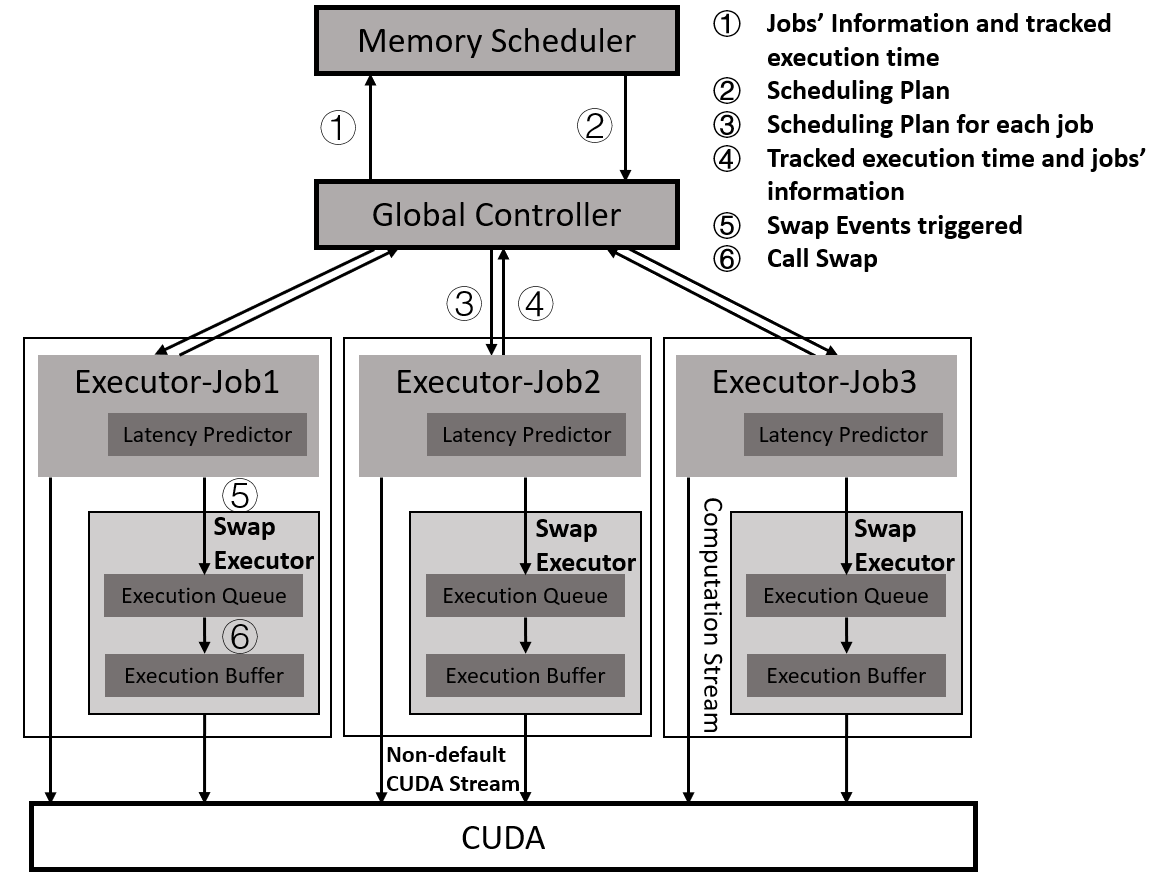}
  \caption{System Architecture, after a job is launched, Executor uses the Latency Predictor to predict the operators' latency and send the information to Memory Scheduler. The Memory scheduler collects all jobs' information and generates a scheduling plan. After receiving the plan from Global Controller, Executor call the Swap Executor to execute the plan.}
  \label{fig:System Architecture}
  \vspace{-0.3cm}
\end{figure}

Since existing approaches cannot schedule among multiple dynamic workloads, we developed a system to support scheduling under such scenarios. The system can collect the necessary information on all the jobs before and during running to support the scheduling algorithm. The information includes the compute graph of jobs, the latency of each operator, and other run-time information such as the GPU utilization. The system will trigger the scheduling algorithm multiple times during startup and running. When receiving the newest scheduling plan, the system will apply it at the next round of iteration. To achieve the functions above, our system contains four components: \textit{Global Controller, Memory Scheduler, Executor, and Swap Executor}, as shown in \autoref{fig:System Architecture}.

\textit{\textbf{Global Controller.}} The \textit{Global Controller} is responsible for launching new jobs' processes and delivering information between the \textit{Executor} and the \textit{Memory Scheduler}. When the user launches a new job, the \textit{Global Controller} creates a new sub-process of the job's \textit{Executor}, and the threads for \textit{Swap Executor}. It is also responsible for communicating with the \textit{Memory Scheduler}, and each job's information is organized and sent to the \textit{Memory Scheduler}. Also, the scheduling plan of each job is distributed to the corresponding \textit{Executor}. With the global controller, our system can support the scheduling algorithm described in \autoref{Introduction}V. 

\textit{\textbf{Memory Scheduler.}} The \textit{Memory Scheduler} takes the jobs' information as inputs, such as the compute graph and the operators' latency tracked by the \textit{Executor}. Such information is used to generate the \textit{Tensor Access Sequence} and the scheduling plan, which consists of multiple swapping, releasing, and recomputation instructions. The detailed algorithm will be described in \autoref{Introduction}V. Each instruction is described by a tuple $(trigger, \Delta time)$, the $trigger$ is the previous operator's end time, and the $\Delta time$ is the time interval after the trigger. Therefore, the corresponding \textit{Swap Event} will be triggered in $\Delta time$ after the $trigger$, just as the \autoref{fig:Concepts' Examples}(c) shows. These instructions are grouped by job id and will be distributed to the corresponding \textit{Executor} by the \textit{Global Controller}.

\begin{figure}[h]
  \centering
  \includegraphics[width=\linewidth]{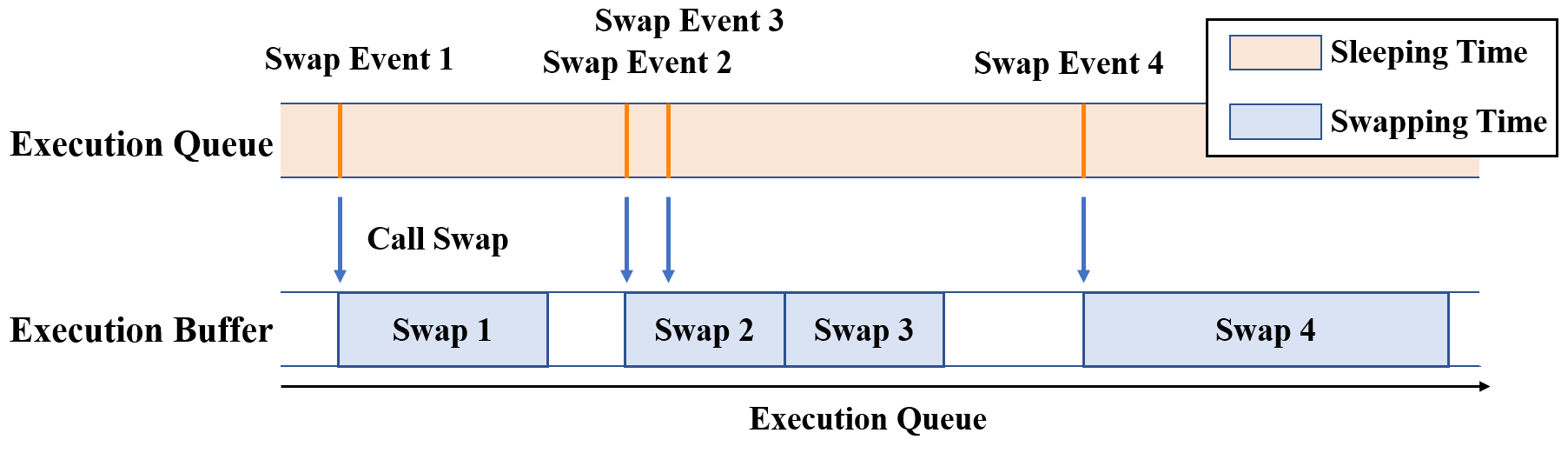}
  \caption{Swap Executor Architecture, Execution Queue tracks the delta time of Swap Event and call the Execution Buffer to swap tensors.}
  \label{fig:Swap Executor Architecture}
\end{figure}

\textit{\textbf{Swap Executor}}. As \autoref{fig:Swap Executor Architecture} shows, the \textit{Swap Executor} contains an Execution Queue and an Execution Buffer. The Execution Queue is a thread with an event queue in it. It receives the swap events triggered by a certain operator (trigger tensor) and pushes these swap events into the queue in chronological order. The Execution Queue pops up the swap events to the Execution Buffer for implementing swapping according to their time intervals. 


\textit{\textbf{Executor.}} The \textit{Executor} executes each operator node in the graph by topological order. For each node, it checks whether all input tensors are in the GPU memory before the computation since the inaccuracy of the estimated \textit{Tensor Access Sequence} may lead to delays in swapping. If that happens, the \textit{Executor} will pause the computation and swap the tensors into the GPU memory. When the computation is finished, certain tensors will be released if the \textit{Memory Scheduler} has set the operator as the release trigger for these tensors. Before releasing, the \textit{Executor} checks the \textit{Swap Executor}. The release and computation processes will be blocked if some \textit{Swap-Out Events} have not been executed. By such a synchronous method, the execution order of swapping, releasing, and computation can be kept. Otherwise, if a given tensor's \textit{Swap-Out Event} is executed later than its release, then the data of the tensor will be lost and would cause exceptions. 

The \textit{Executor} also tracks every operator's latency periodically and sends the information to the \textit{Memory Scheduler} through the \textit{Global Controller} for updating the scheduling plan. The \textit{Executor} will check whether the scheduling plan has been updated each time before a new round of calculation. During the execution of the compute graph, if the scheduling plan requires the current tensor access to trigger some \textit{Swap Events}, the \textit{Executor} will send the \textit{Swap Event} to the Execution Queue of the \textit{Swap Executor} waiting for execution. At the end of an iteration, the Executor pauses to wait for the unfinished swap events of this iteration.

The system can collect information from each job and execute the corresponding scheduling plan with these components. The whole scheduling procedure has four steps:
\begin{enumerate}
    \item Collecting the new jobs information through the \textit{Global Controller}.
    \item The \textit{Memory Scheduler} generates the initial scheduling plan and distributes it to the corresponding job's \textit{Executor} through the \textit{Global Controller}.
    \item The \textit{Swap Executor} executes the scheduling plan during computing. Also, the \textit{Executor} performs the recomputation events at running.
    \item The \textit{Executor} collects the time cost of each operator and reports to the \textit{Global Controller}. When the current operators' latency and the latency used to generate the previous scheduling plan deviate more than the preset threshold, the \textit{Global Controller} will call the \textit{Memory Scheduler} to update the scheduling plan based on the newest estimated \textit{Tensor Access Sequence}.
\end{enumerate}

\section{Methods}
In this section, we detail our algorithms of TENSILE in the order of actual execution, i.e.,
Tensor Access Sequence Generation (\autoref{Tensor Access Sequence Generation}), GPU Memory Peak Analysis(\autoref{GPU memory Peak Analysis}), Swap Event Scheduling\autoref{Swap Event Scheduling}), Recomputation Scheduling(\autoref{Recomputation Scheduling}), and Scheduling Plan Updating(\autoref{Scheduling Plan Updating}).

\subsection{Tensor Access Sequence Generation}
\label{Tensor Access Sequence Generation}
We first introduce our tensor access sequence generation method since our method is based on GPU memory peak analysis, which depends on tensor access sequence generation. We use a learn-based latency prediction method to solve the cold-starting problem mentioned in Section I. In consideration that there are two pieces of information required to infer the \textit{Tensor Access Sequence}, which are the \textit{Tensor Accesses'} order and the latency of each \textit{Tensor Access}. We can deal with this problem by diving into two sub-problems of inferring the \textit{Tensor Access} order and getting each \textit{Tensor Access's} latency. 

As mentioned above, we execute the operators in the topological order with the compute graph. Therefore, we can solve the first sub-problem and get the {Tensor Accesses'} order statically based on each operator's inputs and outputs. It only solves the first sub-problem, which is inferring the access order. The execution order is unrelated to the latency.

However, accurately predicting the CUDA kernel's latency is a challenging problem, especially with multiple dynamic workloads. 


The most straightforward way to solve the second sub-problem is to run the job with 'Passive Mode'. However, it will cause extra overhead since calculation and swapping can not overlap, and the computation must wait for swapping. To make the initial scheduling plan more efficient, our solution uses pre-trained machine learning models to predict each GPU operator's latency with the input data size and the current GPU utilization rates. When the system initializes, it will measure each GPU operator's latency with different input data and GPU usage to generate the training data.

Since the latency of a given operator depends on its inputs' shape, the parameters such as strides of convolution operator, and the current GPU usage, the inputs of our prediction model are denoted as $inputs \equiv <dim_{x_1}^0, ..., dim_{x_1}^m, ...,dim_{x_n}^k, para_1, ..., para_s,gpu\_usage>$, where the $x_1$ to $x_n$ denotes the input tensors of the operator, the $dim_{x_j}^i$ denotes the i-th dimension's size of input $x_j$, the $para_i$ denotes the i-th parameter of the operator, and the $gpu\_usage$ denotes the current GPU's utilization rate. We used a four layers MLP network as the prediction model since the input is not complex, and a simple MLP model is enough to give a relatively precise prediction. According to our experiment, the average MSE of every operator's prediction result is 4.269,  which means that our method can give a relatively precise prediction to most operators. A detailed evaluation of this method's influence on the scheduling process will be introduced in \autoref{Performance of TENSILE's Each Phase}.

Although the GPU's utilization rate is a general representation of the GPU usage condition, and the prediction results will not be precisely accurate, we can keep tracking the operators' latency and dynamically correct the predicted result with the exponential moving weighted average method \cite{EMWA} during the job's run-time. The details will be introduced in \autoref{Scheduling Plan Updating}.

\subsection{GPU Memory Peak Analysis}
\label{GPU memory Peak Analysis}

To reduce the entire system's GPU memory peak, the scheduling algorithm has to know when the GPU memory peak appears and caused by which tensors so that it can try to schedule the \textit{Swap Event} of these tensors. Thus, with the information inferred from the compute graph, we develop an algorithm to analyze the GPU memory peak statistically. 

The algorithm is a discrete simulation process of the \autoref{equation_MFj}. According to the \autoref{equation_MFj}, the GPU memory footprint is changed only when these five situations happen.
\begin{enumerate}
    \item \textit{Iteration Beginning}: At the beginning of a training iteration, these tensors will be in the GPU memory:
    \begin{itemize}
        \item Input data of the model
        \item Parameters which are not swapped out from the last iteration
        \item Initial space used by CUDA, cuDNN, and cuBLAS.
    \end{itemize}
    These tensors will be used to initialize the GPU memory footprint.
    
    \item \textit{Tensor Generating Access}: When a tensor is generated in the GPU memory via $TGA$, GPU memory footprint increases. Note that when updating a parameter, we logically treat the updated parameter as a new tensor, but the new parameter will not cause an increase in GPU memory footprint since it uses the memory address of the old parameter.
    
    \item \textit{Swap-In Event}: A \textit{Swap-In Event} causes GPU memory consumption increase. Since we only concern with the peak of GPU memory footprint, the \textit{Swap-In Event's} finishing time point can be regarded as the time point of the GPU memory footprint increases.
    
    \item \textit{Swap-Out Event}: When a tensor is copied to Host Memory and released, the GPU memory footprint is reduced at the end of the swap-out event (or at the end of the corresponding $TUA$, if the $TUA$ ends later).
    \item \textit{Tensor Release}: For a \textit{Tensor Access $i$} which needs to recompute or swap in, when the last \textit{Tensor Access $i$-1} finished, the tensor can be released from GPU memory, so that the GPU memory footprint will decrease.
\end{enumerate}

Based on the above discussions, we just need to sort the \textit{Tensor Accesses} and \textit{Swap Events} by the time they trigger the GPU memory footprint change. Then we can traverse the job's timeline to determine the GPU memory peak. We then introduce the analysis algorithm in detail as shown in \autoref{algorithm GPU Memory peak Analysis}.


\setlength{\textfloatsep}{0pt}
\begin{algorithm}[htbp]
    \footnotesize
    \caption{GPU Memory Peak Analysis}
    \label{algorithm GPU Memory peak Analysis}
    \LinesNumbered
    \KwIn{$TAS$: tensor access sequence; $SE$: swap events}
    \KwOut{$MP$: memory peak; $MPTensor$: tensors caused memory peak; $LUA$: last input access for each job before memory peak; $MPTime$: time of the memory peak}
    initialize the $memory\_used$ and the $in\_gpu\_tensors$ with the tensors which will be updated in the optimizing phase and the inputs/outputs of the deep learning model.\;
    \For{\textbf{each} $event$ \textbf{in} $time\_axis$}
    {
        $time\longleftarrow event.time$\;
        Release the tensor in $tensors\_to\_be\_release$ if the present time is after its end time\;
        \uIf{$event.type=output\_access$ \textbf{and} $event.tensor$ is not used to initialize the $memory\_used$ and the $in\_gpu\_tensors$}
        {
            $memory\_used \longleftarrow memory\_used + event.size$\;
            $in\_gpu\_tensors.add(event.tensor)$\;
        }
        \uElseIf{$event.type=input\_access$}
        {
            \uIf{$event.release=True$}
            {
                $tensors\_to\_be\_released.append(event)$\;
            }
            \uElse
            {
                $last\_input\_access \longleftarrow event$\;
            }
        }
        \uElseIf{$event.type=swap\_out$ \textbf{or} $event.type=swap\_in$}
        {   
            $last\_access \longleftarrow get\_last\_access(event, time\_axis)$\;
            $event.execute\_ref\longleftarrow last\_access$\;
            $event.execute\_time\longleftarrow last\_access.time$\;
            \uIf{$event.type=swap\_in$}
            {
                $memory\_used \longleftarrow memory\_used + event.size$\;
                $in\_gpu\_tensors.add(event.tensor)$\;
            }
            \uElse
            {
                $memory\_used \longleftarrow memory\_used - event.size$\;
                $in\_gpu\_tensors.remove(event.tensor)$\;
            }
        }
        \uIf{$memory\_used > memory\_peak$}
        {
            $memory\_peak \longleftarrow memory\_used$\;
            $MPTensor \longleftarrow copy(in\_gpu\_tensors)$\;
            $LUA \longleftarrow last\_input\_access$\;
            $MPTime \longleftarrow time$\;
        }
    }
    \textbf{return} $memory\_peak$, $MPTensor$, $LUA$, $MPTime$\;
\end{algorithm}
\vspace{-0.5cm}


\subsection{Swap Event Scheduling}
\label{Swap Event Scheduling}

The primary approach in our system to reduce the GPU memory peak is by swapping tensors out to the Host Memory when they are not in use. It is obvious that swapping every tensor out of the GPU memory after their $TGA$, and swapping them back right before their $TUA$ can save the most GPU memory. However, due to the data transfer occupying the PCIe channels exclusively \cite{Capuchin}, there can only be one tensor being swapped simultaneously. Therefore, swapping tensors within and between tasks must be scheduled appropriately.

\begin{algorithm}[htbp]
    \footnotesize
    \caption{Swap Scheduling Algorithm}
    \label{algorithm Swap Scheduling}
    \LinesNumbered
    \KwIn{$TAS$: tensor access sequence; $SE$: swapped events; $MPT$: the tensors that cause the memory peak; $MSR$: the max swapping rate limit for each job, $SON$: swapped out tensors number for each job; $TAT$: $TAS$ grouped by tensor and sorted by time; all the outputs of \autoref{algorithm GPU Memory peak Analysis}}
    \KwResult{Add new \textit{Swap Events} to $SE$}
    \SetAlgoLined
    \SetKwProg{Fn}{Function}{}{end}
    \Fn{scheduling\_swap($SE$: list, $tensor$):}
    {
        $succeed$, $have\_first\_access$ $\longleftarrow False$\;
        Generate $swap\_out\_event$ and $feasible\_intervals$\;
        \For{\textbf{each} $intervals$ \textbf{in} $feasible\_intervals$}
        {
            \uIf{$intervals$ covers $swap\_out\_event$}
            {
                $SE.add(swap\_out\_event)$\;
                $succeed\_swap\_out \longleftarrow True$\;
                \uIf{$tensor$ \rm is an updated parameter}
                {
                    $first\_access \longleftarrow$ the first $TUA$ of the corresponding non-updated parameter\;
                }
                \uElse
                {
                    $first\_access \longleftarrow$ the first $TUA$ after $access$\;
                }
                $have\_first\_access \longleftarrow $first\_access$ != None$\;
                Generate $swap\_in\_event$ from the $first\_access$\; 
                \uIf{$try\_schedule\_swap\_in(first\_access)$ \rm success}
                {
                    $SE.add(swap\_in\_event)$\;
                    $succeed \longleftarrow True$\;
                }
                \uElse
                {
                    $SE.remove(swap\_out\_event)$\;
                    $swap\_out\_event.latest\_time \longleftarrow first\_access.end\_time$\;
                }
                \textbf{break}\;
            }
        }
        return $succeed$, $succeed\_swap\_out$, $have\_first\_access$\;
    }
    \For{\textbf{each} $tensor$ \textbf{in} $MPT$}
    {
        Infer the $latest\_time$ from the peak analysis result\;
        Infer the $earliest\_time$ from the $TGA$ of the swap-out event\;
        \uIf{tensor\ \rm has\ not\ been\ swapped\ out}
        {
            \uIf{$tensor$ \rm is an updated parameter in optimizing phase}
            {
                scheduling\_swap($SE$, $tensor$)\;
            }
            \uElseIf{SON[tensor.job\_id] $\leq$ MSR[tensor.job\_id] \ \textbf{and}\ TAT[tensor].length$>$1}
            {
                $succeed \longleftarrow False$\;
                $accesses \longleftarrow TAT[tensor]$\;
                $succeed\_swap\_out$, $have\_first\_access$ $\longleftarrow True$\;
                \While{\textbf{not} $succeed$ \textbf{and} $latest\_time > earliest\_time$ \textbf{and} $succeed\_swap\_out$ \textbf{and} $have\_first\_access$}
                {
                    $succeed$, $succeed\_swap\_out$, $have\_first\_access$ $\longleftarrow$ scheduling\_swap($SE$, $tensor$)\;
                }
                \tcp{Ascending sort by end time}
                Try to swap-in the rest of accesses greedily if the above scheduling succeeds\; 
            }
        }
    }
\end{algorithm}

The scheduling process is a deformation of the task scheduling problem with the next two difficulties:

\begin{itemize}
    \item A \textit{Swap Event} cannot be executed without prerequisites. For example, a \textit{Swap-In Event} can be executed if the corresponding \textit{Swap-Out Event} can be performed. Otherwise, no data will be swapped into the GPU memory. Also, to avoid interrupting task execution, the \textit{Swap-Out Event} can be executed only when the tensor's \textit{Swap-In Event} can be scheduled successfully before the next \textit{Tensor Access} (we do not allow passively swapping in for performance reasons).
    \item The scheduling process must be performed considering global jobs' information, and it needs to be updated frequently due to the dynamic workloads. This fact determines that we can not use a very complex algorithm to generate the scheduling plan. Otherwise, the generation of the scheduling plan will cause significantly extra overheads and makes the scheduling plan can not be updated in time under the fluctuation of GPU utilization.
\end{itemize}
Facing the difficulties above, we design a greedy algorithm to balance the performance and the time overhead. The main idea is to swap the largest available tensor whose access will cause the GPU memory footprint peak. Since the scheduling for each tensor can change the GPU memory peak, we must analyze the GPU memory peak and generate the swap events iteratively. In particular, the GPU memory peak will be analyzed first based on the \textit{Tensor Access Sequence}, and choosing the largest tensor that caused the peak to swap. This process is iterated several times until there are no tensors to be scheduled.

With such a greedy algorithm, each scheduled tensor can significantly reduce the GPU memory peak. Now we will detail our algorithm.


We firstly choose the biggest \textit{MPT} among all jobs as the most valuable tensor to swap (candidate tensor). Note that each job has a max swapping rate limit. Considering the swapping among all jobs, the PCIe channel will be jammed if we swap the tensors in each job as many as possible. Also, we can not control the order of swap events among all jobs because jobs in our system are running asynchronously for performance reasons, and each job's time cost of one training step is significantly different. If we use a synchronous way to execute these jobs, then there must be extra synchronization overheads. From this perspective, to reduce the conflict opportunity, we define the max swapping rate limit for each job's tensor swapping as the rate of tensors swapped in a job to tensors swapped in the whole system. It can be regarded as a hyper-parameter. The max swapping rate can be considered the job's priority. Because the tensors' swapping can not always overlap with computation with 100 percent odds due to the conflict on PCIe and the changing GPU utilization, swapping implies the job may run slower than usual. In this perspective, a lower swapping rate means higher priority.

After choosing the candidate tensor, each \textit{Swap Event's} feasible time intervals will be inferred, as shown in \autoref{algorithm Swap Scheduling} Line 4 and Lines 23-24, which is defined as its $earliest\_time$ and $latest\_time$. There are three bases for inferring the feasible time interval:   
\begin{itemize}
    \item The latest time of the \textit{Swap-Out Event} cannot be later than the time the GPU memory peak appears since we want to swap it to reduce the GPU memory peak. Furthermore, a \textit{Swap-Out Event} must start after its $TGA$. 
    \item A \textit{Swap-In Event} cannot be earlier than the corresponding \textit{Swap-Out Event} or later than the $TUA$ that uses it as input. Therefore, our algorithm needs to decide the start time of each event under the constraint of its feasible time intervals. 
    \item The \textit{Swap Event} can not overlap with the corresponding tensor's access. 
\end{itemize}

Based on the feasible time intervals, we choose the available intervals greedily. For a \textit{Swap-Out Event}, it will start as early as possible, and for a \textit{Swap-In Event}, it will start as late as possible so that we can save the GPU memory occupied by the tensor for a longer time. If there are no available intervals, then we will break the outer loop in Line 20 and try to swap the following tensor.

Next, we have to decide when to execute the \textit{Swap-Out Event} and the \textit{Swap-In Event}. Since the tensors in the forward/backward propagation phase and the optimizing phase have different constraints, we will introduce them separately.

The tensors in the forward/backward propagation phase must be swapped into the GPU before the first $TUA$ after the \textit{Swap-Out Event}. Otherwise, an exception will be triggered due to missing necessary input tensors, and the system has to pause the computation to wait for their swapping. As shown in \autoref{algorithm Swap Scheduling} Lines 10-14, we look for the first $TUA$ after the \textit{Swap-Out Event} and generate the \textit{Swap-In Event} and its feasible intervals to see whether any of these intervals can cover the \textit{Swap-In Event}. If they can not cover the event, we will update the \textit{Swap-Out Event}'s earliest start time and regenerate the feasible intervals(Lines 18-19). Otherwise, for the rest $TUA$ of the tensor, we try to swap them in as much as possible(Line 34). If a \textit{Swap Event} is scheduled successfully, the corresponding tensor can be released after the previous $TUA$ is finished. TENSILE will also analyze the compute graph and release the tensors after their last access(Activity Analysis). The detailed algorithm of swapping scheduling is shown in \autoref{algorithm Swap Scheduling}.

Most prior works ignore the tensors which need to be updated in the Opt phase. This causes the 'Across-iteration Scheduling' problem. TENSILE solves this problem with the following rules: if these tensors are swapped out at the optimizing phase, the corresponding tensors must be swapped in at the beginning of the next iteration. As shown in \autoref{algorithm Swap Scheduling} Lines 8-9 and Line 27, it tries to swap out the updated tensors and then swaps in the corresponding tensor before the first $TUA$. \autoref{fig:Across Iteration}(c) shows this process more vividly.

With this algorithm, we can select the most valuable tensors to be appropriately scheduled among all jobs, thus reducing the entire computation job's GPU memory peak.
\vspace{-0.3cm}
\subsection{Recomputation Scheduling}
\label{Recomputation Scheduling}

If the predicted GPU memory peak is still larger than the GPU memory after swapping scheduling, we need to use recomputation to save more GPU memory. However, a major shortcoming of recomputation is that it will add extra $TUA$ and change the \textit{Tensor Access Sequence}. If the inputs of a recomputation have been swapped out or released, then it needs to be swapped in or recomputed, which will disrupt the swapping plan just generated and cause much more time overhead. Therefore, we will only apply recomputation on those tensor accesses which have never been released. To achieve the best GPU memory saving effect, we will select those accesses that conform to the above conditions with the largest recomputation value until the GPU memory is enough to run these jobs.

To measure a recomputation's value, we used the metric proposed in Capuchin \cite{Capuchin}, known as MSPS. This metric can measure the memory saving and extra time cost at the same time.
\begin{equation}
  MSPS \equiv \frac{Memory\ Saving}{Recomputation\ Time}
\end{equation}

\subsection{Scheduling Plan Updating}
\label{Scheduling Plan Updating}
As discussed in \autoref{Introduction}, each operator's latency changes along with the jobs' running in the multiple dynamic workloads scenarios. This leads to a changing \textit{Tensor Access Sequence}, so the initial scheduling plan based on the outdated \textit{Tensor Access Sequence} will be less efficient. For example, the hit rate of \textit{Swap-In Event} will decrease and causes an increasing extra overhead. To adapt to such kinds of dynamic workloads, the \textit{Memory Scheduler} needs to be able to update the scheduling plan based on the operators' latency collected by the \textit{Executor}.

Two major problems should be solved.

\begin{enumerate}
    \item When should the scheduling plan needs to be updated?
    \item How to update the scheduling plan?
\end{enumerate}

Since the reason for updating the scheduling plan is that the latency of operators has been significantly changed compared to the time used to generate the last \textit{Tensor Access Sequence}, we only need to propose a metric to measure this change. We use the average latency rate of change, which is defined as $\frac{|s^{'}-s|}{s}$, where $s^{'}$ is the newest sum of each operation's latency, and $s$ is the last sum of each operation's latency. If the changing rate is larger than the updating threshold set before, then the updating of the scheduling plan will be triggered.

To correct the \textit{Tensor Access Sequence} and make the scheduling plan more accurate, we use the exponential weighted moving average algorithm\cite{EMWA} to update the latency of operators. By predicting and correcting the latency, TENSILE can achieve higher performance than Capuchin's 'Passive Mode' since passively swapping tensors requires extra overhead.

\begin{algorithm}[tbp]
    \footnotesize
    \caption{Complete Scheduling Algorithm}
    \label{algorithm Complete Scheduling}
    \LinesNumbered
    \KwIn{$TAS$: tensor access sequence}
    \SetKwRepeat{Do}{do}{while}
    \textbf{initialization} $swap\_succeed\longleftarrow True$, $recomputation\_succeed\longleftarrow True$, $iter\longleftarrow 0$\;
    run Activity Analysis\;
    \While{$swap\_succeed=True$ \textbf{or} $recomputation\_succeed=True$}
    {
        \uIf{the average $MP_j$ of the jobs corresponding to the selected tensors in the past 3 iterations reduces less than 0.05\% \textbf{and} $iter > 100$}{
            \textbf{break}\;
        }
        \uIf{$iter=0$}{
            \textbf{Run} the $GPU\_Memory\_Peak\_Analysis$ algorithm for each job and merge their GPU memory footprint as $MP$\;
        }
        \uIf{$swap\_succeed=True$}
        {
            $swap\_succeed \longleftarrow False$\;
            \textbf{Run} $Swap\_Scheduling$ to generate swap events\;
            \uIf{no swap events generated}
            {
                $swap\_succeed \longleftarrow False$\;
            }
        }
        \uElseIf{$MP\ge Free\_GPU\_Memory$}
        {
            \textbf{Run} $Recomputation\_Scheduling$\;
            \uIf{no recomputation events generated}
            {
               $recomputation\_succeed \longleftarrow False$\;
            }
        }
        $iter = iter+1$\;
    }
\end{algorithm}
\setlength{\textfloatsep}{0pt}
\subsection{Discussion of supporting Dynamic Workload and the range of application}

With \autoref{algorithm GPU Memory peak Analysis}, TENSILE can give a more effective initial scheduling plan than the 'Passive Mode.' The system will keep tracing recent tensor accesses time to correct the \textit{Tensor Access Sequence} and update the scheduling plan. When a job is launched, the \textit{Tensor Access Sequences} of other jobs are most likely to fluctuate greatly. In this situation, TENSILE updates the \textit{Tensor Access Sequence} based on current GPU usage for all jobs. To avoid extra overhead caused by scheduling plan changes, the system will apply the new plan before computing the next batch of data. The complete scheduling algorithm is shown in \autoref{algorithm Complete Scheduling}. 
Firstly, we run an activity analysis at the beginning of the scheduling to save more memory. Every tensor will be released when its last \textit{TUA} is finished. Then, the \textit{Memory Scheduler} will iteratively run the process below until no swapping or recomputation is scheduled. At each iteration, the algorithm will start with running the \textit{GPU\_Memory\_Peak\_Analysis} function for each job and merge the result as shown in Lines 6-7. Then, if at least one swap event is scheduled successfully in the previous iteration, it will keep trying to swap, as shown in Lines 8-12. Otherwise, if the GPU memory is insufficient, it will try to schedule the recomputation events as shown in Lines 13-15.

Although our method is designed to run on a single GPU, it can also support data-parallel models for multiple GPUs in the cluster. In this situation, the scheduling processes are the same with a single GPU situation on each node, with the premise that its CPU has enough PCIe channels for the GPUs to use in a single node.

\section{Experiments}
To verify the effectiveness of the proposed approaches, we conduct extensive experiments in this section.

\subsection{Methodology}
\textbf{Experiment Platform.} To apply our scheduling algorithm, we implement a naive deep learning framework with CUDA as our experiment platform. The framework is modified based on the tinyflow open-source project\footnote{https://github.com/LB-Yu/tinyflow}. We also reproduce vDNN \cite{vDNN}, and Capuchin \cite{Capuchin} coupling with our framework based on their papers. 

\textbf{Baselines.} We choose two methods as baselines from layer-granularity and tensor-granularity methods to compare with our method. The first one is vDNN, which swaps out the appropriate feature maps and uses a static swap-in strategy for prefetching. There are two types vDNNs, vDNN$_{all}$ and vDNN$_{conv}$, the former tries to swap every feature map, and the latter only swaps the feature maps of convolution layers. We take the vDNN$_{conv}$ as one of our baselines since the authors of vDNN proved that vDNN$_{conv}$ has much higher performance than vDNN$_{all}$ \cite{vDNN}. The second baseline is Capuchin, the newest and the most representative dynamic scheduling method in tensor granularity for a single job. \textbf{Although these baselines are designed for single workload scenarios, we can also create an instance for each workload in our experiments and schedule each workload independently.} And both the baselines and our method overlap data transfer and computation with \textit{cudaMemcpyAsync}.

\textbf{Metrics.} We introduce the metrics used to evaluate the performance of TENSILE. We choose \textbf{Memory Saving Rate(MSR)}, \textbf{Extra Overhead Rate(EOR)}, and \textbf{Cost-Benefit Rate(CBR)} to measure the TENSILE's performance and efficiency compared to the baselines. Their definitions are shown as follows:\\$MSR=\frac{VMP-EMP}{VMP}$; $EOR=\frac{ETC}{VTC}$; $CBR=\frac{MSR}{EOR}$.

Where $VMP$ and $EMP$ are the memory footprint peak of the vanilla group (no scheduling) and the experimental group (TENSILE or baselines), $VTC$ and $ETC$ are the time cost of the vanilla group and the experimental group.

The three metrics we choose are the key metrics to evaluate a GPU memory scheduling method. The Memory Saving Rate (MSR) represents how much GPU memory a scheduling method can save compared to the vanilla method (no scheduling), the higher $MSR$ implies that the corresponding scheduling method can reduce more GPU memory footprint peak compared to the vanilla method. The EOR indicates how many times the time cost will be caused by saving these proportions of GPU memory. The lower $EOR$ implies that the scheduling method has less extra overhead. The CBR is the ratio of MSR and EOR, which indicates how much extra overhead it takes to save a piece of GPU memory. The higher $CBR$ implies that the corresponding scheduling method is more space-time efficient than other methods. These metrics are common and fair for TENSILE and both baselines since we analyzed and compared the MSR, EOR, and CBR of the whole computing process of TENSILE and both baselines in detail. Since TENSILE is designed for multiple dynamic workload scenarios, it is expected to have higher $MSR$ and $CBR$ and less $EOR$ compared to other baseline methods.

\textbf{Statement of our reproduction.} It is necessary to emphasize that the reproduction for Capuchin may not be exactly the same as its authors' implementation in the paper \cite{Capuchin} in the aspect of the system framework since there is no source code to refer to, and it is implemented in TensorFlow. For example, in Capuchin, they measure the GPU kernel latency through CUPTI. However, in our platform, every computation will be synchronized between CPU and GPU, so we can measure the operators' latency in the python codes. We want to compare the scheduling algorithm that runs on the same platform rather than whose platform implementation is more computationally efficient. Therefore, the experiments will be fair since all the methods' scheduling algorithms are implemented exactly as their paper described and run on the same platform for comparison.

\textbf{Experiment Platform.} Our experiment is conducted on a server that has Intel(R) Xeon(R) Silver 4210R CPU, 504GB Host Memory, and one RTX A6000 with 48GB memory. The system version is Ubuntu 18.04 with GCC 7.5.0, the CUDA version is 11.3, the cuDNN is 8.2.0, and Python 3.10.

\textbf{Workloads.} We choose the five neural networks in the evaluation of the Capuchin\cite{Capuchin} to evaluate TENSILE, VGG-16 \cite{VGG}, InceptionV3 \cite{InceptionV3}, InceptionV4 \cite{InceptionV4}, ResNet-50 \cite{ResNet}, and DenseNet \cite{DenseNet}, covered both linear network and non-linear network. We will increase the workload number on one GPU and measure the GPU memory footprint and the extra time overhead for each method.

\textbf{Extra Setting.} The Capuchin is designed to schedule only when the GPU memory is insufficient. It will only save the GPU memory to the threshold that is just able to run the workloads, which makes it difficult to compare its performance with TENSILE since TENSILE is a more active scheduling method that will save GPU memory as much as possible. Therefore, we set Capuchin's GPU memory budget as the same as TENSILE. Then we can compare these two methods' time cost and efficiency fairly. 



Also, our method includes two phases, the cold start phase, and the dynamic updating phase. The former uses predicted operations' latency to generate the schedule plan before the job launching, and the latter updates the scheduling plan with the running information collected from the framework. To analyze these two phases separately, we denote the cold start phase by $\rm TENSILE_{cs}$ and the updating phase by $\rm TENSILE_{up}$. The $\rm TENSILE_{cs}$'s performance is measured once the job is launched, and the $\rm TENSILE_{up}$'s performance is measured once the scheduling plan is updated. We also use $\rm TENSILE$ to denote the whole process of TENSILE. 


\subsection{Performance of TENSILE's Each Phase}
\label{Performance of TENSILE's Each Phase}
We first evaluate the performance of the cold start phase and the update phase of TENSILE by measuring the memory footprint and time cost separately. The results are shown in \autoref{tab:Each Phase's Performance}.

\begin{table}[htbp]
  \caption{Each Phase's Performance}
  \vspace{-0.3cm}
  \label{tab:Each Phase's Performance}
  \begin{center}
   \begin{tabular}{llccc}
    \toprule[1pt]
    Workloads & Phase & MSR & EOR & CBR\\
    \midrule[1pt]
    \multirow{3}*{VGG-16}       & $\rm TENSILE$ & 0.2670 & 1.6287 & 0.1639\\
                                & $\rm TENSILE_{cs}$ & 0.2670 & 6.0128 & 0.0444\\ 
                                & $\rm TENSILE_{up}$ & \textbf{0.3192} & \textbf{1.2475} & \textbf{0.2559}\\
    \midrule[1pt]
    \multirow{3}*{InceptionV3}  & $\rm TENSILE$ & 0.4361 & 1.6468 & 0.2648\\
                                & $\rm TENSILE_{cs}$ & 0.4361 & 3.1128 & 0.1401\\ 
                                & $\rm TENSILE_{up}$ & \textbf{0.4788} & \textbf{1.4839} & \textbf{0.3227}\\
    \midrule[1pt]
    \multirow{3}*{InceptionV4}  & $\rm TENSILE$ & 0.5154 & 1.4281 & 0.3609\\
                                & $\rm TENSILE_{cs}$ & 0.5154 & 2.6727 & 0.1929\\ 
                                & $\rm TENSILE_{up}$ & \textbf{0.5558} & \textbf{1.2898} & \textbf{0.4309}\\
    \midrule[1pt]
    \multirow{3}*{ResNet-50}    & $\rm TENSILE$ & 0.4258 & 1.5540 & 0.2740\\
                                & $\rm TENSILE_{cs}$ & 0.4258 & 3.5395 & 0.1203\\ 
                                & $\rm TENSILE_{up}$ & \textbf{0.4647} & \textbf{1.3813} & \textbf{0.3364}\\
    \midrule[1pt]
    \multirow{3}*{DenseNet}     & $\rm TENSILE$ & 0.5135 & 1.1678 & 0.4397\\
                                & $\rm TENSILE_{cs}$ & 0.5135 & 2.2532 & 0.2279\\ 
                                & $\rm TENSILE_{up}$ & \textbf{0.5236} & \textbf{1.0664} & \textbf{0.4910}\\
    \bottomrule[1pt]
    \end{tabular}
    \end{center}
    \vspace{-0.2cm}
\end{table}

The MSR of $\rm TENSILE$ equals the metric of $\rm TENSILE_{cs}$, which proves that the cold start phase is the bottleneck of the whole process of TENSILE. Also, compared to Capuchin's performance in \autoref{tab:Single Task Performance}, $\rm TENSILE_{cs}$ can achieve a similar performance as Capuchin with the prediction method only. This result proves the effectiveness of our prediction method. Note that the time cost of the cold start phase includes the overhead of the prediction of the MLP models, and this result proves the effectiveness of the scheduling plan updating method. Furthermore, benefiting from the scheduling plan updating method in \autoref{Scheduling Plan Updating}, the update phase's MSR and EOR are significantly higher than the cold start phase. Since the plan updating method takes the main ingredient, its high efficiency significantly reduces the overhead of the entire computing process. 

We also measure the first iteration of TENSILE and Capuchin to compare the EOR of our MLP-based prediction method with the 'Passive Mode' of Capuchin. The results are shown in \autoref{tab:Cold Start Time Cost}, which shows that TENSILE has a more efficient method for the cold-starting situation. For the MLP latency prediction models' memory cost, our method contains 40 models, and the size of their parameters is 321MB.

\begin{table}[htbp]
  \caption{First Iteration Time Cost}
  \vspace{-0.3cm}
  \label{tab:Cold Start Time Cost}
  \begin{center}
  \resizebox{.45\textwidth}{!}{
       \begin{tabular}{llcccc}
        \toprule[1pt]
        Workloads                       & Method             & EOR & Prediction Cost(s) & Prediction EOR \\
        \midrule[1pt]
        \multirow{2}*{VGG-16}           & $\rm TENSILE$      & \textbf{1.2523}   & 1.3549            & 0.2067                    \\
                                        & $\rm Capuchin$     & 5.8053   & -                 & -                     \\ 
        \midrule[1pt]
        \multirow{2}*{InceptionV3}      & $\rm TENSILE$      & \textbf{1.4744}   & 1.4743            & 0.3699                     \\
                                        & $\rm Capuchin$     & 6.7529   & -                 & -                     \\ 
        \midrule[1pt]
        \multirow{2}*{InceptionV4}      & $\rm TENSILE$      & \textbf{1.3797}   & 2.4991            & 0.2865                     \\
                                        & $\rm Capuchin$     & 10.2088  & -                 & -                     \\ 
        \midrule[1pt]
        \multirow{2}*{ResNet-50}        & $\rm TENSILE$      & \textbf{1.3924}  & 1.7823           & 0.3198                     \\
                                        & $\rm Capuchin$     & 9.6073  & -                & -                     \\ 
        \midrule[1pt]
        \multirow{2}*{DenseNet}         & $\rm TENSILE$      & \textbf{1.3855}  & 1.3855           & 0.3335                     \\
                                        & $\rm Capuchin$     & 10.0925 & -                & -                     \\ 
        \bottomrule[1pt]
        \end{tabular}
    }
    \end{center}
    \vspace{-1 cm}
\end{table}

\subsection{Single Workload Performance}
We compare TENSILE's performance with vDNN and Capuchin on a single workload. The batch size is set to 16. And for TENSILE, the max swapping rate limit is set to 100\% for each task since there is only one task in this experiment. The result is shown in \autoref{tab:Single Task Performance}. 




And the result also shows that our method can save 26.70\%-51.54\% memory than the vanilla method, which is much higher than vDNN. This is foreseeable since TENSILE is a tensor granularity scheduling method. And the cost-benefit rate of TENSILE is also much higher than vDNN and Capuchin, which proves the efficiency of TENSILE. 


The high efficiency of TENSILE is due to the fact we put the extra time cost into priority consideration, which is embodied in that we do not allow the non-overlapping between swapping and computation logically. On the contrary, the other baseline methods allow the computation to wait for swapping when scheduling. Also, compared to Capuchin, TENSILE can schedule tensors across iterations. This feature helps to avoid some passive swap-in.

This result proves that although TENSILE is designed for multiple jobs and dynamic workloads, it can still achieve the best efficiency and memory saving rate in a single workload situation. 

\begin{table}[htbp]
  \caption{Single Task Performance}
  \vspace{-0.3cm}
  \label{tab:Single Task Performance}
  \begin{center}
       \begin{tabular}{ccccc}
        \toprule
        Workloads & Methods & MSR & EOR & CBR\\
        \midrule
        \multirow{3}*{VGG-16}
        & vDNN  & 0.1109 & 1.7835 & 0.0622\\
        & Capuchin & 0.2558 & 1.8304 & 0.1398\\
        & $\rm TENSILE$ & \textbf{0.2670} & \textbf{1.6287} & \textbf{0.1639}\\
        \midrule
        \multirow{3}*{InceptionV3}
        & vDNN  & 0.1633 & 2.5184 & 0.0648\\
        & Capuchin & 0.4331 & 2.7745 & 0.1561\\
        & $\rm TENSILE$ & \textbf{0.4361} & \textbf{1.6468} & \textbf{0.2648}\\
        \midrule
        \multirow{3}*{InceptionV4}
        &  vDNN  & 0.1903 & 2.5028 & 0.0760\\
        & Capuchin & 0.5129 & 3.2362 & 0.1585\\
        & $\rm TENSILE$ & \textbf{0.5154} & \textbf{1.4281} & \textbf{0.3609}\\
        \midrule
        \multirow{3}*{ResNet-50}
        & vDNN  & 0.1733 & 2.2981 & 0.0754\\
        & Capuchin & 0.3859 & 2.2320 & 0.1729\\
        & $\rm TENSILE$ & \textbf{0.4258} & \textbf{1.5540} & \textbf{0.2740}\\
        \midrule
        \multirow{3}*{DenseNet}
        & vDNN  & 0.1381 & 2.6148 & 0.0528\\
        & Capuchin & 0.4860 & 3.4400 & 0.1413\\
        & $\rm TENSILE$ & \textbf{0.5135} & \textbf{1.1678} & \textbf{0.4397}\\
        \bottomrule
        \end{tabular}
     \end{center}
\end{table}

\subsection{Scalability}
\begin{figure}[htbp]
  \centering
  \includegraphics[width=\linewidth]{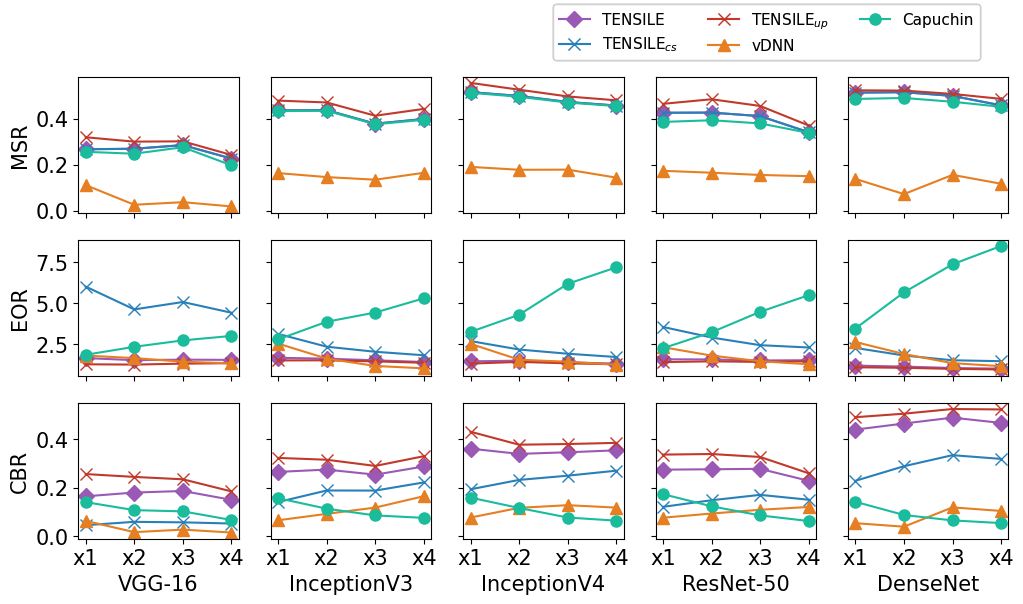}
  \caption{Multiple Dynamic Workloads Performance. TENSILE outperforms both baselines in MSR, EOR, and CBR. And with the increasing workload number, the ratio of TENSILE’s CBR to Capuchin’s CBR keeps increasing in most results. This phenomenon represents that the performance benefits of TENSILE higher than single workload scenario.}
  \label{fig:Multiple Dynamic Workloads Performance}
  \vspace{-0.5cm}
\end{figure}

To evaluate our method's performance on multiple dynamic workloads and test its scalability, we choose one to four workloads respectively and launch them simultaneously to simulate a multiple dynamic workloads scenario. We repeat each test three times and report the average metric in \autoref{fig:Multiple Dynamic Workloads Performance}.

As the \autoref{fig:Multiple Dynamic Workloads Performance} shows, in most of the workloads, TENSILE has achieved the best MSR and CBR, which is solid proof of the performance of TENSILE in multiple dynamic workloads situations. The EOR and CBR of $\rm TENSILE_{cs}$ do not outperform Capuchin in VGG-16 since the VGG-16 has a massive tensor generated by a wide MLP layer, which causes it much harder to schedule without the overhead. TENSILE tackles this problem by the updating method, and the whole TENSILE, including the updating phase, outperforms Capuchin. With the increased workload, TENSILE's MSR of InceptionV3, InceptionV4, ResNet-50, and DenseNet drops slowly. And in all experiments, $\rm TENSILE_{up}$ saves the most GPU memory.

Furthermore, according to the EOR metric, Capuchin's overhead dramatically increases as we analyze in the last paragraph of \autoref{Related Work}. The VDNN's EOR reduces with the increasing workload numbers since the VDNN is a layer-granularity method, which can not schedule all tensors so it has many free time intervals for swapping. This makes its extra time cost less than the vanilla method's time cost. And with increasing workload numbers, the vanilla time cost rises faster than the extra time cost. So the EOR gets lower. Although VDNN has a lower overhead in multiple workload scenarios, its MSR is meager compared to TENSILE and Capuchin.

In any case, the efficiency of TENSILE is still much higher than both baselines. Moreover, the ratios of TENSILE's CBR to vDNN/Capuchin's CBR under the multi-workload environment are higher than in the single-workload environment. The conclusion is that TENSILE can run these workloads with less (or at least the same) GPU memory and time cost than baselines. Especially compared to the Capuchin, whose EOR is significantly increasing when the number of workloads increases, which implies most of their swapping transfer can not overlap with the computation, TENSILES's EOR is much more stable. This experiment strongly proves that the performance of TENSILE is much better than vDNN and Capuchin in multiple dynamic workloads scenarios. To evaluate our method's performance on multiple dynamic workloads and test its scalability, we choose one to four workloads respectively and launch them simultaneously to simulate a multiple dynamic workloads scenario. We repeat each test three times and report the average metric in \autoref{fig:Multiple Dynamic Workloads Performance}.

\subsection{Mixed Neural Architectures}
To measure the TENSILE's performance in a mixed neural architectures workloads scenario, we randomly launch the five above workloads and measure the total GPU memory footprint and execution overhead. The five workloads are launched one by one in random order, and this procedure is repeated three times. We set the budget of each workload of Capuchin with the MSR of TENSILE in \autoref{tab:Mixed Neural Architectures Workloads Performance}.

As \autoref{tab:Mixed Neural Architectures Workloads Performance} shows, TENSILE is comprehensively leading in EOR and CBR under different neural architectures. It can save much more GPU memory with less extra cost in scenarios with complex non-linear workloads, making it more suitable for environments such as in-database machine learning.

\subsection{Overhead Analysis}
Although TENSILE outperforms two baselines, the overheads during GPU memory scheduling can hardly be avoided completely.

For single workloads scenarios, the overheads are caused by two reasons. The first reason is that when the memory is extremely insufficient, the overhead of passive swap-in and recomputation cannot be avoided completely. However, using a more efficient scheduling algorithm, TENSILE achieves less overhead than Capuchin. The second reason is the failed tensor prefetch caused by the fluctuation of GPU operator latency and the error of estimated operator latency.

For multiple workloads scenarios, the GPU utilization rate and the PCIe channel are all influenced by the asynchronous workloads, which makes the prefetch much harder and makes more passive swap-in and recomputation. TENSILE alleviates this problem by choosing the most valuable tensors among all workloads to schedule to achieve higher efficiency. However, it is not a perfect solution, and we will continue to find better solutions to this problem in the future.



%

\begin{table}[htbp]
  \caption{Mixed Neural Architectures Workloads Performance}
  \vspace{-0.3cm}
  \label{tab:Mixed Neural Architectures Workloads Performance}
  \begin{center}
  \begin{tabular}{cccc}
    \toprule
    Methods & MSR & EOR & CBR\\
    \midrule
    vDNN  & 0.0580 & 1.6297 & 0.0356\\
    Capuchin & 0.1150 & 7.6545 & 0.0150\\
    TENSILE & \textbf{0.1565} & \textbf{1.1589} & \textbf{0.1350}\\
    \bottomrule
    \end{tabular}
    \end{center}
\end{table}

\subsection{Influence of Batch Size}
\begin{figure}[htbp]
  \centering
  \includegraphics[width=\linewidth]{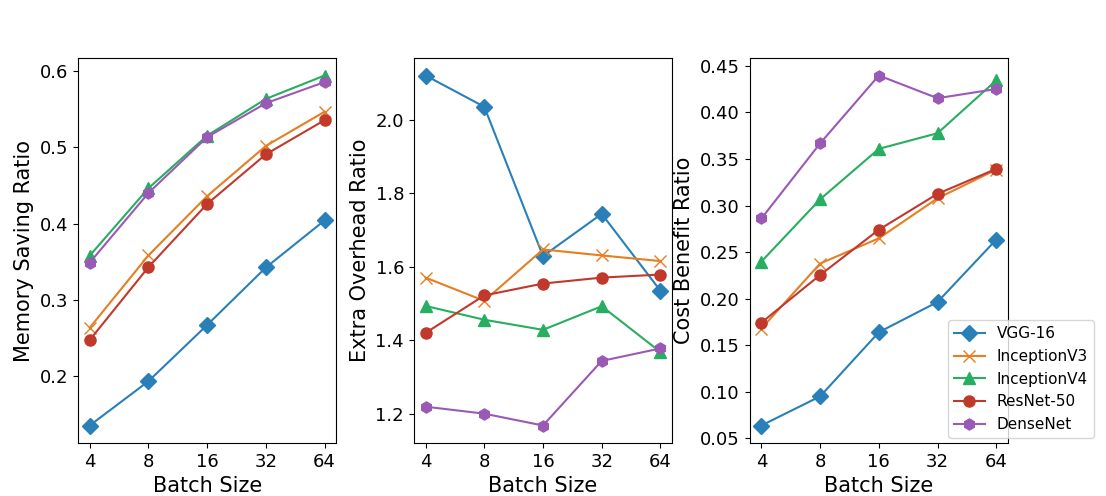}
  \caption{Batch size's influence}
  \label{fig:Batch size's influence}
\end{figure}

We also analyzed the batch size's influence on our method by running those five workloads, respectively, with batch sizes 4, 8, 16, 32, and 64. The results are shown in \autoref{fig:Batch size's influence}.

The MSR increases with the batch size for all workloads, and so does their CBR in general. This result shows that the TENSILE can save more GPU memory with less overhead when the batch size increases. This result proves that TENSILE is more effective in scheduling the interim results than the parameters since the former is the only one influenced by the batch size. The DenseNet with batch size 32 and 64 is an exception. We believe it is because there are too many tensors to swap. And when the batch size increases, the swap events can more easily conflict with each other. And for VGG-16, it is the only workload whose EOR drops with the increase in batch size. This is because that VGG-16 has a wide MLP layer that outputs a huge tensor. This tensor's computation cost increases faster than its scheduling cost, so the EOR gets lower.

\vspace{-0.5cm}

\section{Conclusions \& Future Work}
We propose TENSILE, a tensor granularity GPU memory scheduling method toward multiple dynamic workload scenarios. TENSILE has solved three major problems of prior works, multiple dynamic workloads, cold-starting, and across-iteration scheduling by designing a scheduling method and corresponding system, a latency prediction method, and a scheduling algorithm. Furthermore, TENSILE is a generic scheduling method for computing tasks described as DAGs, especially deep learning tasks. The experiment results show that TENSILE can achieve much higher performance in such scenarios than vDNN and Capuchin. Even in single workload scenarios, TENSILE can still outperform baselines. These features can help systems such as in-database machine learning run more deep learning tasks with tiny extra overhead.

\section*{Acknowledgment}
This paper was supported by NSFC grant (62232005).


%





\newpage
\bibliography{reference}
\bibliographystyle{IEEEtran}

\end{document}